# Hybrid attention structure preserving network for reconstruction of under-sampled OCT images

Zezhao Guo[1], Zhanfang Zhao[1,*]

[1]*College of Information and Engineering, Hebei GEO University, Hebei, China*

*\*zhaozhanfang@hgu.edu.cn*

Optical coherence tomography (OCT) is a non-invasive, high-resolution imaging technology that provides cross-sectional images of tissues. Dense acquisition of A-scans along the fast axis is required to obtain high digital resolution images. However, the dense acquisition will increase the acquisition time, causing the discomfort of patients. In addition, the longer acquisition time may lead to motion artifacts, thereby reducing imaging quality. In this work, we proposed a hybrid attention structure preserving network (HASPN) to achieve super-resolution of under-sampled OCT images to speed up the acquisition. It utilized adaptive dilated convolution-based channel attention (ADCCA) and enhanced spatial attention (ESA) to better capture the channel and spatial information of the feature. Moreover, convolutional neural networks (CNNs) exhibit a higher sensitivity of low-frequency than high-frequency information, which may lead to a limited performance on reconstructing fine structures. To address this problem, we introduced an additional branch, i.e., textures & details branch, using high-frequency decomposition images to better super-resolve retinal structures. The superiority of our method was demonstrated by qualitative and quantitative comparisons with mainstream methods. HASPN was applied to the diabetic macular edema retinal dataset, validating its good generalization ability.

## 1. Introduction

Optical coherence tomography (OCT) is a non-invasive optical imaging technique [1]. Due to its cellular-level imaging resolution, it has been widely used in ophthalmology, dermatology, and cardiology [2-4].

Typically, dense acquisition is required to capture fine microstructures of the sample. However, conducting dense acquisition, especially over a large field of view, will decrease the imaging speed and thereby cause the discomfort of patients. Moreover, the longer acquisition time is likely to exacerbate eye motion, introducing artifacts into the image [5]. Down-sampling is the easiest way to speed up the acquisition, however, at the sacrifice of the resolution.

To improve the digital resolution of under-sampled images, various conventional methods have been proposed. Fang et al. proposed a sparsity-based framework that simultaneously performed interpolation and denoising to reconstruct the OCT images efficiently [6]. Abbasi et al. introduced a non-local weighted sparse representation (NWSR) method to integrate sparse representations of multiple noisy and denoised patches, improving the quality [7]. Wang et al. proposed to utilize compressive sensing (CS) and digital filters to enhance the down-sampled OCT angiography images [8]. The study demonstrated that the vascular structures could be well reconstructed through CS with a sampling rate on B-scans at 70%, suggesting that CS could significantly accelerate acquisition in the OCT system. However, the reconstruction performance of these conventional methods was limited.

In recent years, deep learning methods have been popular among various medical image processing tasks [9-11]. Huang et al. utilized a generative adversarial network (GAN) to super-resolve OCT images while reducing the noise, introducing deep learning into OCT super-resolution for the first time [12]. Qiu et al. proposed a novel semi-supervised method using UNet and DBPN to achieve simultaneous super-resolution and denoising [13]. However, these deep-learning-based super-resolution networks ignore the fact that convolutional neural network (CNN) is more sensitive to low-frequency information [14], potentially limiting the performance on reconstructing fine-grained structures in OCT images.

To obtain high digital resolution images within a short acquisition time, we proposed a novel OCT super-resolution model named hybrid attention structure preserving network (HASPN). HASPN has two branches. One branch was used to primarily restore the low-frequency features of images. The other branch could enhance the perceptual quality of the output by learning the high-frequency features of decomposed images. The low-frequency and high-frequency features from the two branches were concatenated over channels to fuse the information. Additionally, the hybrid attention mechanism was introduced to enhance the network's capacity to learn spatial and channel information, improving the reconstruction capability. Next, we utilized the public retinal OCT image dataset OCT2017 to test HASPN at different sampling rates. Compared with the current mainstream methods, HASPN achieved the best results at 4x and 8x SR. Moreover, we investigated the impact of network depths and widths on performance. Finally, the experiment demonstrated our proposed HASPN exhibited a good generalization ability on the diabetic macular edema (DME) dataset which was unseen during training.

## 2. Methods

### 2.1 Data preparation

In this paper, we utilized the retinal OCT image dataset OCT2017 [15]. The original dataset contains 84,495 images in total, covering normal and abnormal retinal images. From the subset of retinal images, 1,300 images were selected from the subset of normal retinal images as the training set, 200 images for the validation set, and 100 images for the testing set.

Considering the limited GPU resources, the images were randomly cropped into 256x256 as the high-resolution (HR) ground truth. To generate low-resolution (LR) images, we under-sampled the columns of the ground truth, obtaining 2x, 4x, and 8x images. Subsequently, the LR-HR image pairs were obtained. In addition, 100 DME OCT images from OCT2017 dataset were used to generate a dataset for validating the generalization capability of the network.

## 2.2 Image decomposition

Unsharp masking (USM) is commonly employed in image processing to enhance high-frequency details [16]. To be specific, a blurred version of the image is subtracted from the original image to generate a residual image. This residual image is then added back to the original image to enhance edges and details. Its specific steps are as follows:

$$R = O - B, \qquad (1)$$

$$S = O + k(R), \qquad (2)$$

where $O$, $B$, $R$ represent the original image, the blurred image, and the residual image (high-frequency image). $k$ is the scaling coefficient used to adjust the degree of sharpening while $S$ is the sharpened image.

Xu et al. demonstrated that CNN exhibited a greater sensitivity to low-frequency information than high-frequency information [14]. However, high-frequency information is essential for reconstructing fine details. To address this problem, we proposed an approach inspired by USM that involved decomposing images into a residual image enriched with high-frequency content. This residual image was subsequently input into the additional branch to enhance the reconstruction of high-frequency features. Specifically, a Gaussian filter with a kernel size of 5x5 and a kernel standard deviation of 1.5 in the X direction was applied to blur the image. Then, the high-frequency image could be obtained by subtraction. Both LR and HR images were decomposed to generate the corresponding high-frequency images. Different from the Eq. 2, we utilized a textures & details CNN branch to enhance edges and details. The outputs of the original branch and the textures & details branch will then be concatenated and fused to generate the final image.

## 2.3 Hybrid attention mechanism

Previous studies have proven the effectiveness of attention mechanisms in super-resolution tasks [17]. It can enable the network to focus on important features, thereby enhancing the quality of reconstruction. As shown in Fig. 1, we designed a hybrid attention mechanism, i.e., intra-block and inter-block attention. First, we integrated an enhanced spatial attention (ESA) [18] in the spatial attention residual block (SARB). Initially, a 1x1 convolutional layer was performed to reduce the channel dimension, thereby decreasing the computational complexity of the ESA module. And then, a 3x3 convolutional layer with a stride of 2 was utilized to reduce the resolution of the feature map by half. Next, a 7x7 max pooling with a stride of 3 was used to achieve downsampling and enlarge the receptive field. Subsequently, a 3x3 convolutional layer was used for feature extraction. Bilinear and 1x1 Conv were utilized to recover the spatial and channel dimensions, respectively. Finally, the input of ESA was dotted with the attention score matrix. Different from the conventional ESA, we did not use Conv Group (two Conv(3x3)-ReLU and one Conv(3x3)) to extract features. We experimentally found that using a 3x3 convolution for feature extraction was better than using the Conv Group in the original ESA. Specifically, utilizing 3x3 Conv for feature extraction resulted in a superior PSNR, with an increase of 1.87dB, and SSIM, with a 0.017 higher value, compared to utilizing Conv Group. That means adopting a single 3x3 convolution layer not only significantly reduces the model's computational complexity but also achieves better performance. In conclusion, ESA enables the network to focus on specific spatial regions of feature maps, thus enhancing the efficiency of feature extraction.

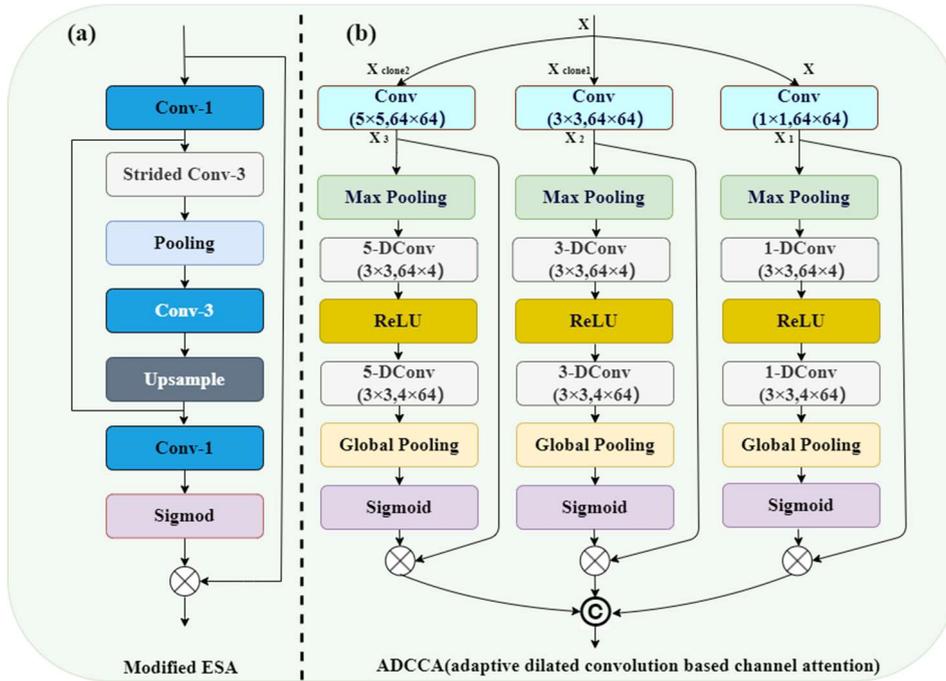

Figure 1. Hybrid attention mechanism in HASPN. (a) Modified ESA, where Conv-N represents a NxN convolutional layer. (b) ADCCA, where N-DConv (kxk, ixo) denotes a kxk convolutional layer with a dilation factor of N, input channel of i, and output channel of o.

Secondly, to further enhance the ability of the network to distinguish the importance of different channels, adaptive dilated convolution-based channel attention (ADCCA) was incorporated every M SARBs (in Fig. 2). ADCCA exploits kernels of different sizes and different dilated factors (1, 3, 5) to enrich the receptive field of convolution, thereby capturing information of various scales. Similar to SENet [19], ADCCA first performs a squeeze operation using max pooling to reduce the resolution of feature maps by half. Then, it follows with an excitation step, which involves dilated Conv-ReLU-dilated Conv (DC-ReLU-DC) operations with different dilated factors. Next, a global pooling is used to decrease the resolution of feature maps to 1x1 to obtain the attention of each channel.

## 2.4 Super resolution network framework

Inspired by the TDPN [20], we proposed a novel network named HASPN as shown in Fig. 2. The network contained two parallel branches: one branch was responsible for restoring the coarse image, while the other branch focused on the restoration of fine textures and details. The outputs of the two branches were finally integrated through a fusion module to generate super-resolution images.

Each branch consisted of three parts: shallow feature extraction, deep feature extraction, and upsampling reconstruction. In the shallow feature extraction stage, a 3x3 convolutional layer was used to extract the shallow features of the network. These features were then fed to each hybrid attention residual block group (HARBG) module through the skip connections from the shared source to help the network better focus on high-frequency features. Deep feature extraction consisted of multiple hybrid attention residual block (HARB) units and a 3x3 convolutional layer. Each HARB unit consisted of M SARBs, an ADCCA, and a feature fusion module (FFM). ESA was introduced into SARB to allow the network focus on some important spatial features, significantly enhancing the perceptual quality of reconstructed images in super-resolution. After processed by multiple SARB units, multi-scale information was extracted by ADCCA which adaptively used convolutions of various sizes and different dilation factors. In addition, it made the network focus more on key feature channels effectively. Finally, FFM was used to merge feature maps at various scales in ADCCA.

Bilinear was utilized as the horizontal upsampling method. The reconstruction module contained a 3x3 Conv, a SARB, and another 3x3 Conv. Finally, the coarse image reconstructed by the original LR image branch and the high-frequency image reconstructed by the textures & details branch were concatenated by channel and fused together. The fusion module has a similar structure to the FFM, with one key difference: it adds a 2-DConv between two convolutional layers. This modification enables the fusion module to capture a larger receptive field and integrate a wider range of contextual information within both branches. As a result, the reconstructed images have richer details and more accurate structures.

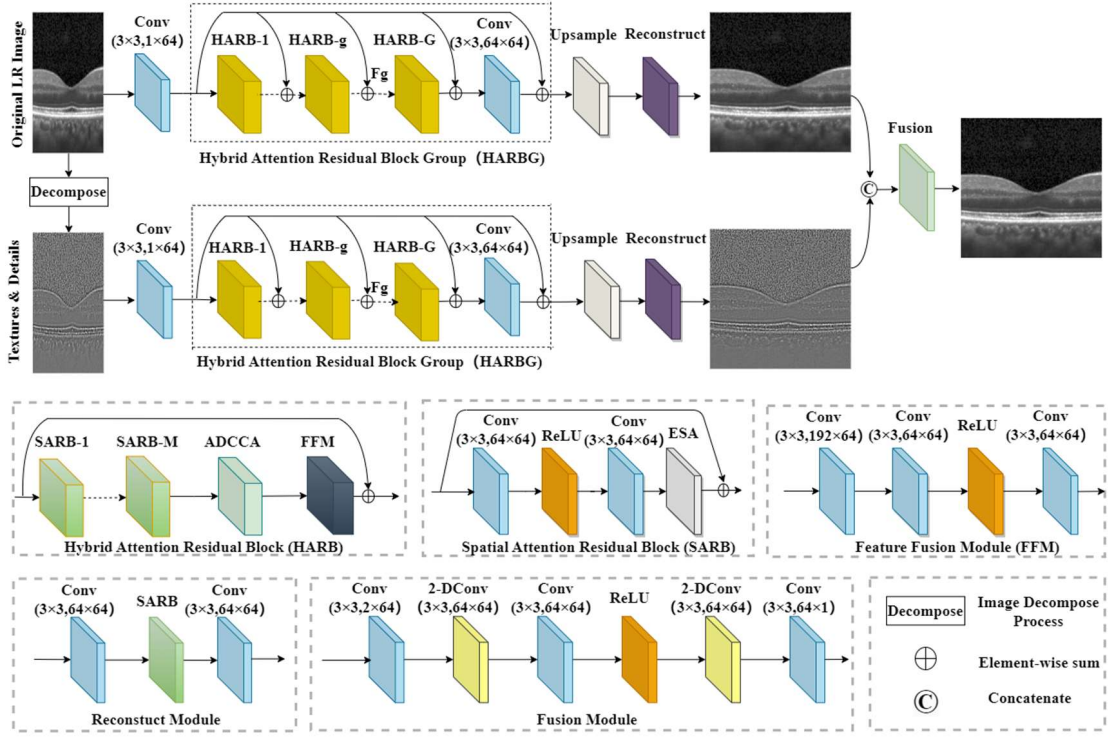

Figure 2. Framework of the proposed HASPN.

## 2.5 Evaluation metrics

To evaluate the performance of the proposed method, two image quality metrics are introduced: peak signal-to-noise ratio (PSNR) [21], structural similarity index metric (SSIM) [22]. PSNR is a commonly used metric to measure the quality of image reconstruction. It evaluates the similarity between the reconstructed image and the original image at the pixel intensity level. It is defined as follows:

$$MSE = \frac{1}{N} \sum_{i=1}^{N} [I_{SR}(i) - I_{HR}(i)]^2, \tag{3}$$

$$PSNR = 10 \log_{10}(\frac{\max(I_{SR})^2}{MSE}), \tag{4}$$

where $I_{SR}$ represents the image reconstructed by the network, and $I_{HR}$ is the ground truth image.

SSIM focuses on the perceptual structure of the image and assess the similarity of images in terms of luminance, contrast, and structure. The definition of SSIM is given by:

$$L(I_{SR}, I_{HR}) = \frac{2\mu_{I_{SR}}\mu_{I_{HR}} + C_1}{\mu_{I_{SR}}^2 + \mu_{I_{HR}}^2 + C_1}, \tag{5}$$

$$C(I_{SR}, I_{HR}) = \frac{2\sigma_{I_{SR}}\sigma_{I_{HR}} + C_2}{\sigma_{I_{SR}}^2 + \sigma_{I_{HR}}^2 + C_2}, \tag{6}$$

$$S(I_{SR}, I_{HR}) = \frac{\sigma_{I_{SR}I_{HR}} + C_3}{\sigma_{I_{SR}}\sigma_{I_{HR}} + C_3}, \tag{7}$$

where $L$, $C$, and $S$ represent luminance, contrast, and structure, respectively. $\mu_{I_{SR}}$, $\mu_{I_{HR}}$ are the mean of $I_{SR}$ and $I_{HR}$, respectively. While $\sigma_{I_{SR}}$, $\sigma_{I_{HR}}$ are the variance of $I_{SR}$ and $I_{HR}$, respectively. $\sigma_{I_{SR}I_{HR}}$ is the covariance of $I_{SR}$ and $I_{HR}$. SSIM is the product of these three components $L$, $C$, and $S$. When $C_3$ is set to $C_2/2$, the final SSIM is as follows:

$$SSIM = \frac{(2\mu_{I_{SR}}\mu_{I_{HR}} + C_1)(2\sigma_{I_{SR}I_{HR}} + C_2)}{(\mu_{I_{SR}}^2 + \mu_{I_{HR}}^2 + C_1)(\sigma_{I_{SR}}^2 + \sigma_{I_{HR}}^2 + C_2)}. \tag{8}$$

## 2.6 Loss function

To pursue high PSNR while preserving more accurate retinal structures, our loss function was defined as:

$$L = L_\alpha + L_\beta + L_\gamma, \tag{9}$$

where $L_\alpha$, $L_\beta$, and $L_\gamma$ represent the losses between the reconstructed coarse image and the ground truth, the reconstructed high-frequency image and the high-frequency image of the ground truth, and the reconstructed image and the ground truth, respectively. $L_\alpha$, $L_\beta$, and $L_\gamma$ were the same function as follows:

$$L_\alpha = L_{pix} + L_{per} + L_{gra}. \tag{10}$$

Lim et al. found that while minimizing L2 norm can maximize the PSNR value, using L1 norm can lead to a better network convergence [23]. Consequently, L1 norm was employed to measure the pixel error between the output and ground truth. $L_{pix}$ was defined as follows:

$$L_{pix} = \frac{1}{N}\sum_{i=1}^{N}\left\|I_{SR}^{i} - I_{HR}^{i}\right\|_1, \tag{11}$$

where $I_{SR}^i$ and $I_{HR}^i$ represent the i-th SR image and i-th HR image in a batch, respectively.

However, only using $L_{pix}$ may not achieve a good perceptual performance. Hence, a perceptual loss [24] was included to enhance visual similarity of the output images to HR images. Specifically, it utilized a pre-trained VGG19 network [25] to extract high-level information at the L-th layer, and employed L2 norm to measure the error of extracted features. $L_{per}$ was defined as follows:

$$L_{per} = \frac{1}{N}\sum_{i=1}^{N}\left\|\Phi^L(I_{SR}^i) - \Phi^L(I_{HR}^i)\right\|_2^2, \tag{12}$$

where $\Phi^L(I_{SR}^i)$ and $\Phi^L(I_{HR}^i)$ represent the features of the i-th SR image extracted by the L-th layer and the features of the i-th HR image extracted by the L-th layer in a batch, respectively.

To avoid the smoothing effect caused by minimizing $L_{pix}$, the gradient loss was used to penalize the gradient of images. $L_{gra}$ was defined as follows:

$$L_{gra} = \frac{1}{N} \sum_{i=1}^{N} \left\| \nabla I_{SR}^i - \nabla I_{HR}^i \right\|_1, \quad (13)$$

where $\nabla I_{SR}^i$ and $\nabla I_{HR}^i$ represent the gradient operator of the i-th SR image and the gradient operator of the i-th HR image in a batch, respectively. $\nabla I_{SR}^i$ was defined as follows:

$$\nabla I_{SR}^i = \frac{\partial^2 I_{SR}^i(x,y)}{\partial x^2} + \frac{\partial^2 I_{SR}^i(x,y)}{\partial y^2}, \quad (14)$$

$$\frac{\partial^2 I_{SR}^i(x,y)}{\partial x^2} = I_{SR}^i(x+1,y) + I_{SR}^i(x-1,y) - 2I_{SR}^i(x,y), \quad (15)$$

$$\frac{\partial^2 I_{SR}^i(x,y)}{\partial y^2} = I_{SR}^i(x,y+1) + I_{SR}^i(x,y-1) - 2I_{SR}^i(x,y). \quad (16)$$

## 2.7 Implementation details

During training, the hyperparameters G and M for HASPN were set to 20 and 5, respectively. All the networks were optimized using the Adam optimizer with $\beta_1$=0.9 and $\beta_2$=0.999, with a initial learning rate of 1e-4. The learning rate for each layer across all networks decayed by 50% every 20 epochs. The batch size for each network were 2. All models were trained for 200 epochs to ensure their convergences.

The entire process was implemented within the PyTorch 2.1.0 framework, compatible with Python version 3.10, on the Tesla A100 GPU with 40GB.

## 3. Results and discussion

To demonstrate the superiority of our proposed network HASPN, it was qualitatively compared with prevailing methods, including Bicubic, SRCNN [26], FSRCNN [27], EDSR [23], RDN [28], RCAN [29], SRGAN [30], ESRGAN [31], RFANet [18], TDPN [20].

As shown in Fig. 3 (pair_72), the outer segment (OS) of the Bicubic reconstructed image exhibited discontinuity. Due to the characteristics of interpolation methods, many ringing artifacts existed at the edges of the external limiting membrane (ELM) and retinal pigment epithelium (RPE). The ELM in the FSRCNN reconstructed image was excessively blurred and affected by artifacts when using deconvolution as the upsampling method [32]. These artifacts significantly affected the final quality of images. Surprisingly, compared to the results of Bicubic and FSRCNN, the ELM reconstructed by SRCNN displayed higher contrast and sharpness. However, the RPE layers in these reconstructed images were severely distorted compared to the HR image. EDSR,

RDN, and RCAN employed a wide-channel residual block, dense residual connections, and channel attention in network designs, respectively, to enhance the network's ability to learn features. This led to better visual performance than Bicubic, SRCNN, and FSRCNN. Additionally, EDSR, RDN, and RCAN achieved results comparable to GAN-based methods (SRGAN, ESRGAN) and RFANet. Furthermore, TDPN reconstructed clearer retinal structures than all methods except our model. However, it is worth noting that TDPN failed to reconstruct the tiny granular structure observed in the RPE layer. In comparison, our model HASPN could reconstruct these subtle structures better, and the restored ELM had higher contrast than other methods. Moreover, the OS reconstructed by HASPN was more continuous.

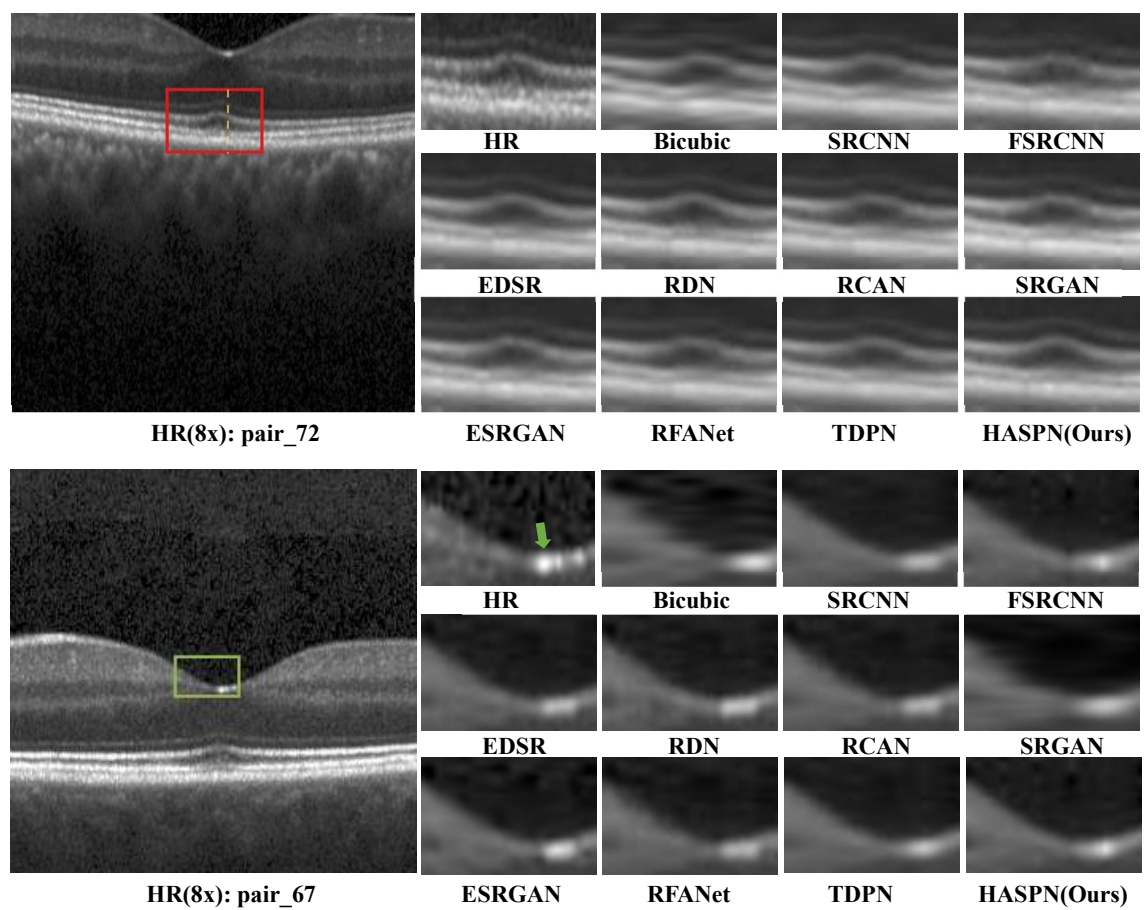

Figure 3. Visual comparisons of HASPN with prevailing models at 8x SR.

For pair_67, the internal limiting membrane (ILM) reconstructed by Bicubic, SRGAN, and RFANet exhibited a ladder-like structure. For the reconstruction of the central fovea (denoted by the green arrow), all methods except FSRCNN and HASPN had large differences with the HR image. RDN and ESRGAN failed to reconstruct the inner nuclear layer (INL). Although the differences between TDPN and the HR image in the reconstruction of INL was smaller, the reconstructed ILM and the central fovea were still slightly blurred. In contrast, our model HASPN shown excellent performance in restoring these structures. It not only completely reconstructed the INL, but also had the biggest visual similarity with the HR image in the ILM. The results demonstrate the high accuracy and superiority of HASPN in reconstructing fine structures in retinal images.

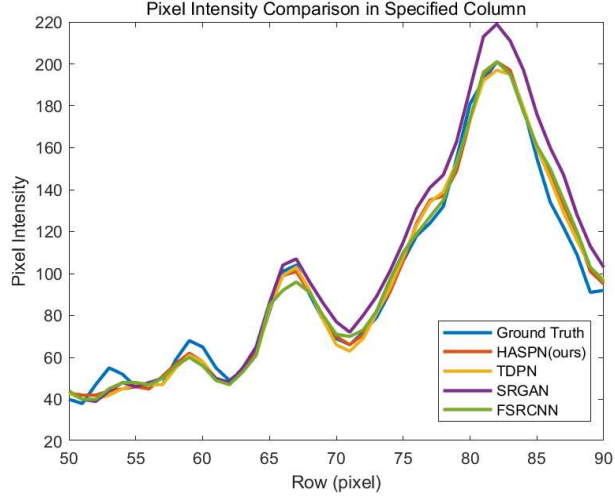

Figure 4. Profile of the orange dashed line in HR(8x): pair_72 of Fig. 3.

Table 1. Quantitative comparisons of 2x, 4x, and 8x SR. The upward arrow (↑) indicates that higher values yield better performance. The best and second best results were highlighted and underlined, respectively.

| Methods | 2x | | 4x | | 8x | |
|---|---|---|---|---|---|---|
| | **PSNR↑** | **SSIM↑** | **PSNR↑** | **SSIM↑** | **PSNR↑** | **SSIM↑** |
| Bicubic | 31.73 | 0.8410 | 27.68 | 0.6633 | 25.74 | 0.5686 |
| SRCNN | 33.62 | 0.8877 | 29.44 | 0.7102 | 27.79 | 0.6276 |
| FSRCNN | 33.66 | 0.8884 | <u>30.11</u> | <u>0.7637</u> | 28.25 | <u>0.6733</u> |
| EDSR | 33.59 | 0.8871 | 29.95 | 0.7593 | 28.29 | 0.6538 |
| RDN | 33.67 | 0.8897 | 29.92 | 0.7589 | 28.27 | 0.6541 |
| RCAN | 32.55 | 0.8667 | 30.01 | 0.7507 | 28.14 | 0.6421` |
| SRGAN | 32.42 | 0.8620 | 29.68 | 0.7402 | <u>28.35</u> | 0.6624 |
| ESRGAN | **33.73** | **0.8913** | 29.74 | 0.7557 | 28.34 | 0.6645 |
| RFANet | <u>33.69</u> | <u>0.8898</u> | 29.97 | 0.7604 | 28.04 | 0.6392 |
| TDPN | 31.36 | 0.8761 | 30.06 | 0.7629 | 27.64 | 0.6482 |
| HASPN(ours) | 32.65 | 0.8881 | **30.14** | **0.7650** | **28.55** | **0.6786** |

Next, to further reflect the performance of our method, we compared HASPN with TDPN, SRGAN (rank second in terms of PNSR), and FSRCNN (rank second in terms of SSIM) by plotting the profile of the selected A-line (indicated by the dashed orange line in HR(8x): pair_72 of Fig. 3). The comparisons were shown in Fig. 4. The peak (the rows from 65 to 70) in Fig. 4 corresponds to the inner/outer segment junction (IS/OS junction) in Fig. 3. It was obvious the structures reconstructed by HASPN and TDPN were very close to the ground truth, while there was a significant disparity between FSRCNN and the ground truth. In particular, the OS (the rows from 70 to 75) reconstructed by HASPN almost overlapped with the ground truth. In contrast, there were noticeable gaps between the OS reconstructed by TDPN, SRGAN, FSRCNN and the ground truth. This demonstrates that HASPN is superior to other methods on preserving retinal layers. Furthermore, within the RPE layer (the rows from 80 to 85), the reconstruction outcomes of HASPN and FSRCNN closely approximated the ground truth. However, the results of TDPN and SRGAN had a distance from the ground truth. This may be because these two methods produce overly smooth edges during the reconstruction.

In Table 1, we quantitatively compared HASPN with other prevailing methods using PSNR and SSIM. It was obvious ESRGAN and RFANet achieved the best and second-best performances respectively at 2x SR. Although HASPN did not achieve the best PSNR at 2x SR, it achieved a SSIM which was only slightly lower than ESRGAN. Surprisingly, FSRCNN ranked second in reconstruction performance at 4x and 8x SR. However, its visual performance was inferior to many prevailing methods. On the other hand, HASPN can achieve the highest PSNR and SSIM at 4x and 8x SR. When super-resolving images at 8x, HASPN exhibited a 0.91 dB higher PSNR compared to TDPN, along with an SSIM that was 0.0304 higher than that of TDPN. Hence, it can be concluded that HASPN can be applied to scenarios where fewer A-scans were acquired.

Table 2. Quantitative comparison of HASPN architectures with different widths and depths at 4x SR. G, M, C represent the number of HARB, SARB, and channel in each layer, respectively.

| G | M | C | PSNR↑ | SSIM↑ |
|---|---|---|---|---|
| **20** | **5** | **64** | **30.14** | **0.7650** |
| 16 | 5 | 64 | 29.57 | 0.7564 |
| 8 | 5 | 64 | 28.66 | 0.7474 |
| 20 | 4 | 64 | 29.83 | 0.7546 |
| 20 | 2 | 64 | 29.17 | 0.7536 |
| 20 | 5 | 32 | 29.12 | 0.7352 |
| 20 | 5 | 16 | 29.08 | 0.7287 |

To investigate the impact of different network depths and widths on performance, we reduced the numbers of HARB (G), SARB (M), and channel (C) respectively. Table 2 indicates that as the numbers of G, M, and C increased, the performance of the model improved. This demonstrates that augmenting the depth and width of the network can effectively enhance its feature extraction and representation capabilities. Specifically, when the numbers of G, M, and C reached 20, 5, and 64, the model achieved the highest PSNR and SSIM at 4x SR. In addition, we observed that G significantly enhanced both PSNR and SSIM at smaller values (8 and 16). However, at smaller values of M (2 and 4), the improvement in SSIM was relatively modest, and at smaller values of C (16 and 32), the increase in PSNR was minimal.

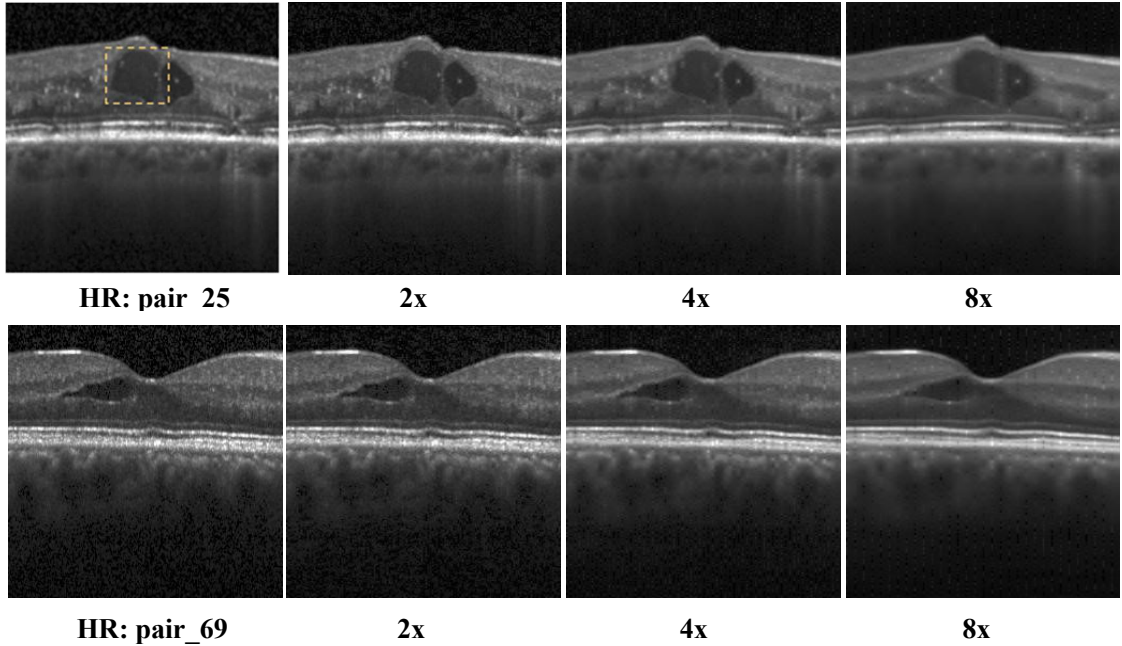

Figure 5. Generalization performance of the proposed network on diabetic macular edema (DME).

Finally, we tested the generalization capability of our proposed model trained with the normal retinal dataset using the DME dataset as mentioned in Section 2.1. As shown in Fig. 5, HASPN can effectively reconstruct structures of the retinal layers at 2x, 4x, and 8x SR. When the upscaling factor was 2x, the reconstructed image was almost same with the HR image. However, when performing super-resolution at 8x, the hyperreflective dots in the intraretinal fluid of the reconstructed image (denoted by the dashed orange rectangle) were not reconstructed well. Except few fine details, our method can reconstruct most of the retinal layer structures. Therefore, it can be concluded that HASPN exhibits an excellent generalization ability and has the potential to be applied into clinical utilization.

## 4. Conclusions

In this work, we proposed a novel hybrid attention structure preserving network (HASPN) to speed up the acquisition while obtaining high digital resolution images comparable to those by dense acquisition. HASPN displays a superior lateral super-resolution reconstruction performance compared to many mainstream super-resolution methods on the public OCT retinal dataset OCT2017. Through qualitative and quantitative analysis, we demonstrated that HASPN could effectively preserve the structural information of OCT under-sampled images and restore more details. Moreover, HASPN achieved the best results at 4x and 8x SR. It proves that HASPN can be applied to some scenes that need to acquire fewer A-scans. In addition, we investigated the impact of depths and widths on the performance of the network. Finally, we validated that HASPN had an excellent generalization capability and could be applied to reconstruct cross-domain OCT images. Our future research will explore self-supervised methods for reconstructing under-sampled OCT images. Additionally, we will consider applying HASPN to other medical imaging modalities such as magnetic resonance imaging and computed tomography to expand its use in medical research and applications.

**Data availability statement**.

The code of this study is available at https://github.com/ZacharyG666/HASPN-for-OCT.

**Acknowledgments.**

This work was supported by Hebei Provincial Social Science Foundation Project (No. HB20TQ003).